\documentclass[prl,twocolumn]{revtex4}
\usepackage{graphicx}

\begin{document}

\title{Nanopositioning of a diamond nanocrystal containing a single NV defect center}

\author{T. van der Sar$^1$, E.C. Heeres$^2$, G.M. Dmochowski$^1$, G. de Lange $^1$, L. Robledo$^1$, T.H. Oosterkamp$^2$, and R. Hanson$^1$}
\affiliation{{$^1$Kavli Institute of Nanoscience, Delft University of Technology, P.O. Box 5046, 2600 GA Delft, The Netherlands}
\newline{$^2$Leiden Institute of Physics, Leiden University, Niels Bohrweg 2, 2333 CA Leiden, The Netherlands}}

\date{\today}

\begin{abstract}
Precise control over the position of a single quantum object is important for many experiments in quantum science and nanotechnology. We report on a technique for high-accuracy positioning of individual diamond nanocrystals. The positioning is done with a home-built nanomanipulator under real-time scanning electron imaging, yielding an accuracy of a few nanometers. This technique is applied to pick up, move and position a single NV defect center contained in a diamond nanocrystal. We verify that the unique optical and spin properties of the NV center are conserved by the positioning process.
\end{abstract}

\maketitle

The coupling of a single quantum object to degrees of freedom in its environment is a central theme in quantum science and engineering. Examples are the coupling of a single photon source to an optical resonator~\cite{Vahala03,Kimble08} or to a plasmonic waveguide~\cite{Akimov07}, and the coupling of a single spin to surrounding spins~\cite{Childress06,Hanson08a}. Studying and engineering such couplings is not only of fundamental interest, but may also lead to dramatic improvements in fluorescence detection efficiency~\cite{Chang07}, ultrasensitive magnetometry ~\cite{Chernobrod05,Degen08,Taylor08,Maze08,Bala08} and applications in quantum information processing~\cite{Kimble08, Hanson08b,Neumann08}. Controlled and precise positioning of the quantum object under study is essential for many of these experiments.

Examples of well-studied single photon emitters in a solid-state environment are quantum dots, fluorescing dye molecules and nitrogen-vacancy (NV) centers in diamond. In particular NV centers, which consist of a substitutional nitrogen atom next to an adjacent vacancy in the diamond lattice, are extremely stable sources of single photons~\cite{Kurtsiefer00,Beveratos01}. In addition, they have the unique property of a paramagnetic spin whose quantum state can be read out optically using fluorescence microscopy~\cite{Gruber97} and be coherently controlled using magnetic resonance~\cite{Jelezko04a}. Crucially, all these properties are retained under ambient conditions. Since NV centers can form in nanocrystals as small as 10~nm~\cite{Boersch09}, their position can in principle be controlled with an accuracy of a few nanometers.

The potential of NV centers for quantum optical experiments is underlined by recent reports demonstrating coupling to confined optical modes of a microsphere~\cite{Park06,Schietinger08} and a microdisk cavity~\cite{Barclay08}. In these experiments, however, the positioning accuracy of the diamond nanocrystals was determined by the resolution of the optical setup ($\sim$500nm), whereas subwavelength resolution is desired.

\begin{figure}[b]
\begin{center}
\includegraphics[width=8.5cm]{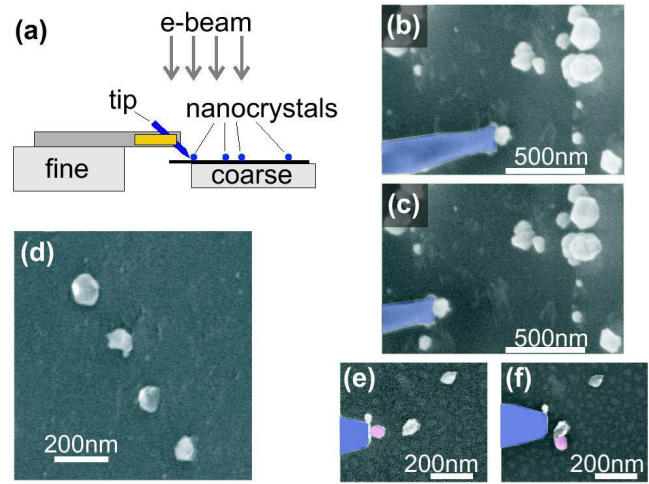}
\end{center}
\vspace*{-.5cm}
\caption{(color online)(a) Schematic of the nanomanipulator. The coarse stage can be moved in three dimensions by Attocube piezo steppers (range 3~mm, smallest step size 25~nm). A sharp, chemically etched tungsten tip is mounted on the piezo fine stage (range 15~$\mu$m, smallest step size sub-nm) which can also be moved in three dimensions. Both stages are controlled by an analog joystick, allowing for accurate adjustment of their positions. (b)-(c) SEM images showing a nanocrystal being attached to the tip (shaded blue) (b) and subsequently being lift from the substrate by retraction of the tip (c). (d) SEM image of four diamond nanocrystals that were subsequently picked up and positioned to form a line. (e)-(f) SEM images showing a nanocrystal (shaded pink) first attached to the tip (e), then being positioned onto another nanocrystal (f).}
\vspace*{-.3cm}
\label{fig:nanomanipulator}
\end{figure}

Here, we demonstrate a versatile technique to position a single NV center contained in a diamond nanocrystal to an arbitrary location. We locate and characterize diamond nanocrystals with single NV centers using a scanning confocal microscope. Subsequently, we use a home-built nanomanipulator, consisting of a sharp probe mounted on a piezo-electrically controlled system inside a scanning electron microscope (SEM), for picking up and positioning the nanocrystal with nanometer precision. This technique is directly applicable to studies of the coupling of a single photon emitter in a nanocrystal to various photonic and plasmonic structures, and can facilitate experiments in magnetometry and quantum information processing.

First, we demonstrate our ability to position individual nanocrystals on an empty flat substrate with high spatial accuracy. Then we describe how we locate and identify nanocrystals containing a single NV center. Finally, we demonstrate explicitly through optical characaterization and coherent spin manipulation that the unique properties of the NV center are preserved by the positioning process.

Positioning of nanocrystals is performed inside a FEI Nova NanoSEM using a home-built nanomanipulator located inside the SEM main chamber. The nanomanipulator consists of a coarse stage onto which a sample is mounted, and a fine stage holding a sharp tip that can be used to pick up nanometer-size objects (Fig.~\ref{fig:nanomanipulator}(a))~\cite{Katan08}. In contrast to manipulation of nano-objects with an AFM~\cite{Wang07,Barth08}, this setup allows real-time imaging of the probe with nanometer resolution during the manipulation process.

In Fig.~\ref{fig:nanomanipulator}(b)-(d) we demonstrate our ability to pick up, move and accurately position individual diamond nanocrystals. The crystals (mean size $\sim$100~nm) are dispersed onto an oxidized silicon substrate.  To pick up a nanocrystal, it is first brought within 15~$\mu$m of the tip by moving the coarse stage. Subsequently, control is switched to the fine stage and the tip is translated and lowered until it touches the nanocrystal (Fig.~\ref{fig:nanomanipulator}(b)). The crystal is lifted from the substrate by pushing against it with the tip until it sticks (presumably by van der Waals forces), and then retracting the tip (Fig.~\ref{fig:nanomanipulator}(c)). At this point, the coarse stage can be moved to position the crystal in a new location on the same substrate or on another substrate within the 3~mm range of the stage. Alternatively, substrates can be exchanged while the crystal remains attached to the tip.

\begin{figure}[tb]
\begin{center}
\includegraphics[width=8.5cm]{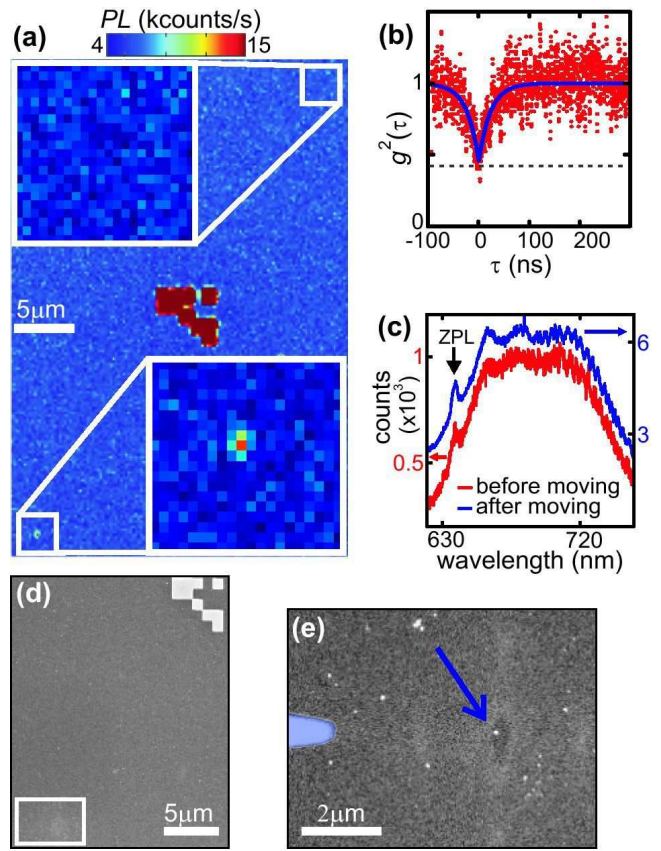}
\end{center}
\vspace*{-.5cm}
\caption{(color online) Identifying and positioning a single NV center. (a) Photoluminescence ($PL$) image of part of the sample around a gold marker. The bright spot in the lowerleft corner is verified to be a single NV center in (b)-(c). (b) Auto-correlation function $g^{(2)}(\tau)$ of photon detection events, showing an antibunching dip below 0.5~\cite{Kurtsiefer00}. The solid line is a fit based on a 3-level model~\cite{Kurtsiefer00}. The dotted line indicates the independently measured contribution of background fluorescence. (c) $PL$ spectrum of the NV center showing the characteristic zero-phonon line (ZPL) around 637~nm~\cite{Gruber97}. Red (blue) data is taken before (after) positioning~\cite{note}. (d) SEM image of the same area as in (a). (e) Zoom-in on the white box in (d), with the arrow pointing to the nanocrystal containing the NV center. The faint white spots next to it are markings made by focussing the laser onto a spot for 2 minutes at a power of 10~mW to facilitate identification of the NV-containing nanocrystal. The contrast difference is presumably due to local oxidation or evaporation of the Cr film. The tip is visible on the left.}
\vspace*{-.3cm}
\label{fig:NV1BeforeMoving}
\end{figure}

The crystal is positioned by following the reverse process. While the initial placement has a precision of $\sim$20~nm (Fig.~\ref{fig:nanomanipulator}(d)), the positioning can be optimized further by pushing the crystal with the tip, which is only limited by the SEM imaging resolution ($\sim$nm). Alternatively, we can position a crystal such that it is touching an on-chip structure such as a nanowire, or another diamond nanocrystal, as is shown in Fig.~\ref{fig:nanomanipulator}(e)-(f).

Next, we demonstrate the positioning of a single NV center. Diamond nanocrystals ($\sim$10-100~nm in size) are dispersed onto a glass cover slip that has been coated with a thin ($\sim$3nm) layer of Cr (to allow for SEM imaging while retaining high optical transparancy) and which contains gold markers (defined by e-beam lithography) for position reference.

NV centers are imaged by fluorescence microscopy using a home-built scanning confocal microscope~\cite{Gruber97}. All data is taken at room temperature. In Fig.~\ref{fig:NV1BeforeMoving}(a), a photoluminescence ($PL$) image of a 30~$\mu$m~x~45~$\mu$m area of the sample is shown. The fluorescing spot in the lowerleft corner is verified to be a single emitter by time correlation measurements on single-photon detection events~\cite{Kurtsiefer00}, see Fig.~\ref{fig:NV1BeforeMoving}(b). The subsequent observation of the characteristic zero-phonon line (ZPL) around 637~nm (Fig.~\ref{fig:NV1BeforeMoving}(c)) shows that this emitter is an NV center~\cite{Gruber97}.

In the SEM, the nanocrystal containing the single NV-center is identified using the reference markers (Fig. \ref{fig:NV1BeforeMoving}d), and subsequently picked up and positioned on the other side of the marker. Although this area contains other nanocrystals, we know from the optical characterization (inset Fig.~\ref{fig:NV1BeforeMoving}(a)) that they do not contain fluorescing defects. Also, the energy of the imaging electrons (10~keV) is far too low to displace carbon atoms~\cite{Koike92} and thus no additional NV centers are formed.

\begin{figure}[tb]
\begin{center}
\includegraphics[width=8.5cm]{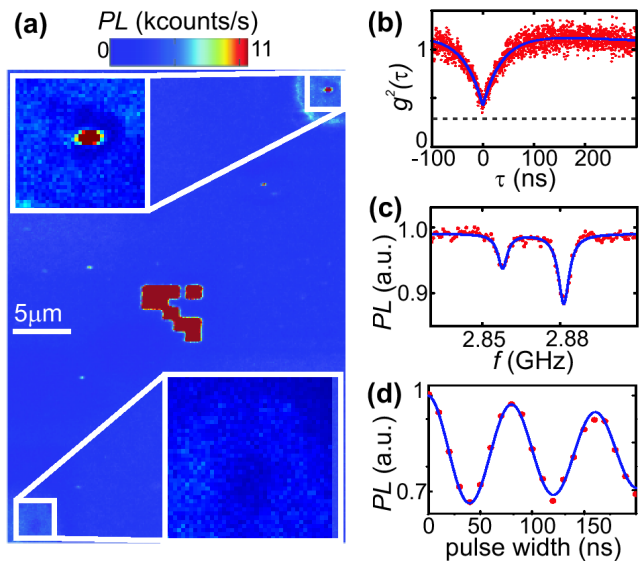}
\end{center}
\vspace*{-.5cm}
\caption{(color online) Optical and spin characterization of a single NV center after positioning. (a) $PL$ image of the same area as in Fig.~\ref{fig:NV1BeforeMoving}(a)~\cite{note,carbon}. The NV-containing crystal has been moved from the lowerleft to the upperright corner. (b) Measurement of $g^2(\tau)$ as in \ref{fig:NV1BeforeMoving}(b), showing an antibunching dip below 0.5. (c) Optically detected magnetic resonance (ODMR) and (d) coherent oscillations of the NV center electron spin. The ODMR splitting is due to a stray magnetic field ($\sim$5~G). Microwaves are applied through a 30~$\mu$m diameter wire that is located in close proximity to the NV center.}
\label{fig:NV1AfterMoving}
\end{figure}

After the nanopositioning, a spatial $PL$ map of the same area as in Fig.~\ref{fig:NV1BeforeMoving}(a) is made (Fig.~\ref{fig:NV1AfterMoving}(a)). To ensure that the NV center was not damaged by the positioning, we repeat the optical characterization. The observation of photon antibunching (Fig.~\ref{fig:NV1AfterMoving}(b)) and the ZPL in the spectrum (Fig.~\ref{fig:NV1BeforeMoving}(c)) confirm that the nanocrystal still contains one NV center.

To explicitly show that we have positioned a single spin, we demonstrate coherent control of the quantum state of the positioned spin. Optically detected magnetic resonance reveals the characteristic zero-field splitting of 2.87~GHz of the NV center~\cite{Gruber97} (Fig.~\ref{fig:NV1AfterMoving}(c)). By using short microwave pulses that are resonant with the spin transition, we observe coherent spin rotations (Rabi nutations) of this single NV center~\cite{Jelezko04a,Hanson08b} (Fig.~\ref{fig:NV1AfterMoving}(d)).

In summary, we have demonstrated a technique to pick up, move, and position individual diamond nanocrystals with few-nanometer precision, all under real-time SEM imaging. This technique has been used to position a single NV center contained in a diamond nanocrystal. We have demonstrated that the unique spin and optical properties of the NV center electron spin are preserved in the positioning process. This technique is directly applicable to controlled studies of the coupling of a single NV center to photonic and plasmonic structures.

We thank  C. Degen, G. Fuchs, M. Hesselberth, F. Jelezko, T. Schenkel and D. van der Zalm, for discussions and experimental help. This work is supported by the Stichting voor Fundamenteel Onderzoek der Materie (FOM) and the Nederlandse Organisatie voor Wetenschappelijk Onderzoek
(NWO). T.H.O. acknowledges support from Technologiestichting STW and from an ERC Starting Grant.


\begin{thebibliography}{99}
\bibitem{note} Note that optical measurements before and after positioning were done with a different microscope objective and therefore do not allow direct quantitative comparison.
\end{thebibliography}
\end{document}